\documentclass[12pt]{article}
\usepackage{amsmath,amssymb}
\usepackage{graphicx,hyperref}
\usepackage[textwidth=18cm,textheight=24.9cm]{geometry}
\begin{document}
\topmargin -2.8cm 
\begin{center}
{\bfseries QCD phase transition studied by means of hadron production 
} 
\vskip 5mm
Jan Rafelski and Michal Petran 
\vskip 5mm
{\small \it Department of Physics, The University of Arizona, Tucson, AZ 85721, USA}
\end{center}
\vskip 5mm

\begin{abstract} 
This is a brief review of our work describing the hadronization process of a QGP fireball formed in relativistic heavy-ion collisions. We introduce the SHARE method of analysis of hadron multiplicities. Using this tool we describe in consistent continuos manner the yield of all hadrons produced in the available range of reaction energies and centralities. The properties of the fireball final state can be understood by considering all primary hadronic particles. The dense hadron fireball created at SPS, RHIC, and LHC shows that the final state is differentiated solely by: i) volume changes; and ii) flavor (strangeness, charm) content. Conversely, emerging particles add up to create universal hadronization pressure $P = 80 \pm 3$ MeV/fm$^3$ for all considered collision systems. The relative strangeness to entropy content of a large fireball is found to be that of quark-gluon plasma degrees of freedom near the chemical QGP equilibrium. This 'Universal Hadronization' condition common to SPS, RHIC, and LHC agrees with the proposed reaction picture of a direct QGP fireball evaporation into free-streaming hadrons. \\[0.3cm]
Presented in Wroclaw at the February 2014 MB32-Symposium honoring Ludwik Turko.\\ 
Appeared in: {\it Physics of Particles and Nuclei} {\bf 46} (5) pp 748-755 (2015)\\ 
\url{https://doi.org/10.1134/S1063779615050238}.

\end{abstract}

\section{Statistical hadronization with resonances (SHARE)
}
The focus of the statistical hadronization model is on particle abundances, i.e. the integrated $p_\perp$ spectra as measured in heavy-ion collision experiments. Our interest is mainly in properties of the source which are evaluated independent of the complex transverse dynamics. This is the reason to analyze the integrated $p_\perp$ spectra. Particle yields allow the exploration of the source properties in the frame comoving with the particles; the collective transverse matter dynamics gets integrated out. 

We describe particle yields within Fermi's statistical approach using Hagedorn's canonical reformulation we call statistical hadronization model (SHM). To wit: by assuming equal hadron production strength irrespective of produced hadron type, the particle yields depend only on the available phase space:
\begin{itemize}
\item 
Fermi Micro-canonical phase space:\\ 
has sharp energy and a sharp number of particles \cite{Fermi:1950jd}. However, since experiments report event-averaged rapidity particle abundances, the model should describe an average event.
\item 
Canonical phase space: \\
has a sharp number of particles, but an ensemble average of energy $E$ which is adjusted by the (inverse) temperature $T$ as a Lagrange multiplier which may be, but needs not be a kinetic process temperature.
\item 
Grand-canonical ensemble phase space:\\ fixes both energy $E$ and number of particles $N$ on average. $N$ is a constraint implemented by the Lagrange multiplier $\mu$, the chemical potential, which is equivalent to the use of the fugacity $\Upsilon = {\rm e}^{\mu/T}$. 
\end{itemize}
We have implemented the SHM in a publicly available program to fit the SHM parameters. The program is called SHARE (= Statistical HAdronization with REsonances) and was released in its first version by Torrieri et al. \cite{Torrieri:2004zz}, then augmented in its second version by fluctuations \cite{Torrieri:2006xi} and in its recently updated version charm was also included \cite{Petran:2013dva}.

SHARE incorporates in its many thousand lines of code the mass spectrum of more than 500 hadrons according to the particle data group (PDG 2012) \cite{Beringer:1900zz}; hadron decays in more than 2500 channels (PDG 2012); integrated hadron yields, ratios and decay cascades. Its output provides the yields of all (presently $\sim 30$) experimentally observed hadron species, and the physical properties of the particle source at hadronization. Bulk matter constraints such as charge per baryon (for heaviest ions $Q/B\sim 0.39$), and that the net strangeness vanishes $\langle s - \bar{s}\rangle=0$ are implemented.

In Fig.~\ref{Fig1} we show the schematics of the SHARE program structure. The fitting of the SHM parameters to observational data proceeds according to the following steps:
\begin{figure}[!ht]
\centering
\includegraphics[width=0.85\textwidth]{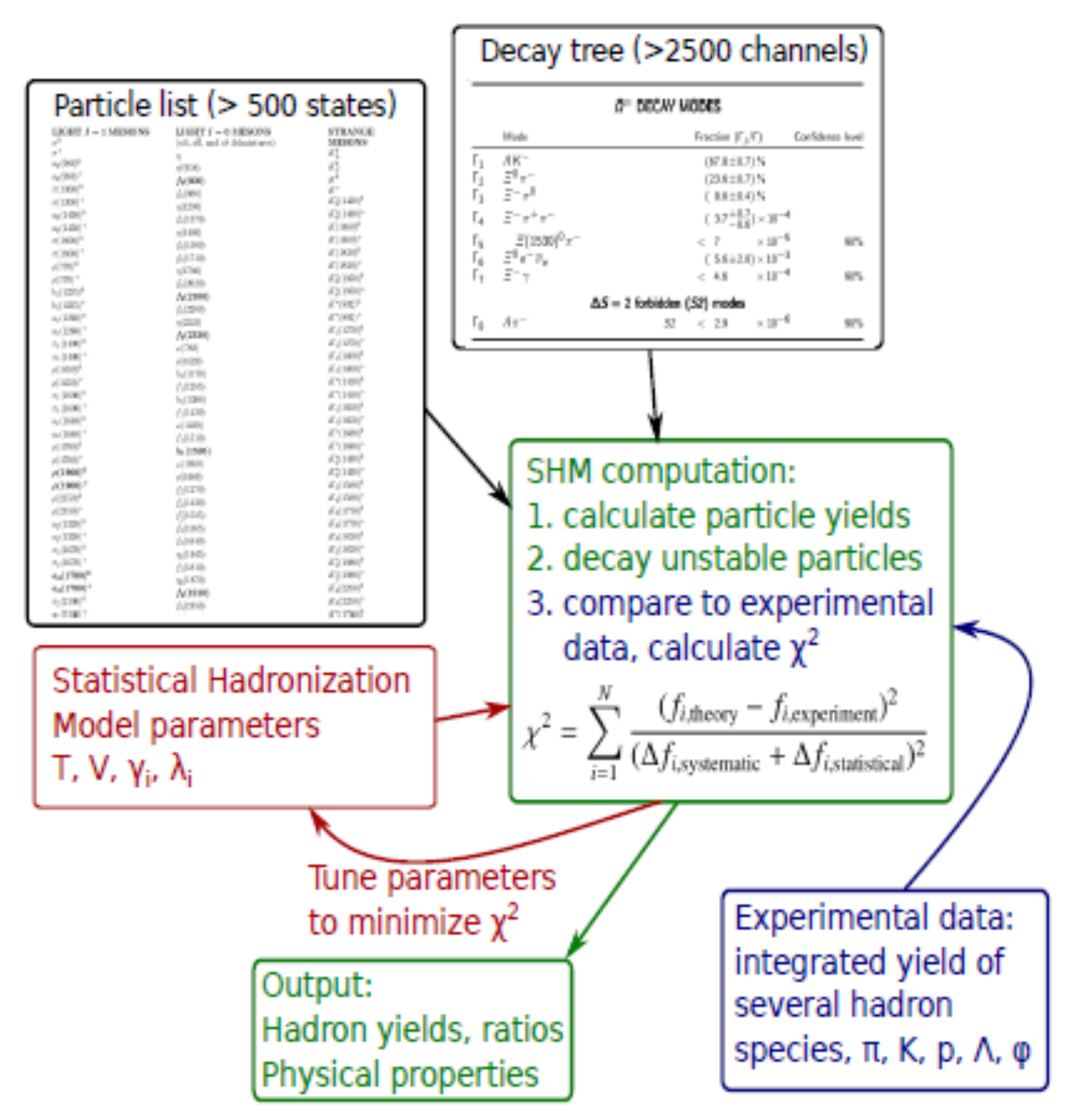}
\caption{Schematics of the SHARE program structure.}
\label{Fig1}
\end{figure}
\begin{enumerate}
\item Input by hypothesis: $T$, $V$, $\gamma_q$, $\gamma_s$, $\lambda_q$, $\lambda_s$, $\lambda_3$
\item Compute the yields of all primary hadrons
\item Account for decay feed-down to observed particles
\item Evaluate bulk properties and bulk constraints
\item Compare to experimental data and evaluate $\chi^2$ including bulk constraints
\item Use $\chi^2$ minimization strategies to tune parameters to match data and constraints -- with new parameters go back to item 1.
\end{enumerate} 
SHARE iterates these steps till CERN provided programs for parameter optimization terminates. Several initial input parameters sets can be tried to assure that the same best, stable parameter fit is found. If such a solution was not achieved, it is advisable to evolve a better initial input parameter set from fits that worked nearby either in energy or centrality. 
In order to account for the quark flavor chemistry we introduce the following quantities which go back to the initial model of QGP hadronization of 30 years ago \cite{Koch:1986ud}, for illustration see also Fig.~\ref{Fig2}:
\begin{figure}[!ht]
\centering
\includegraphics[width=6cm]{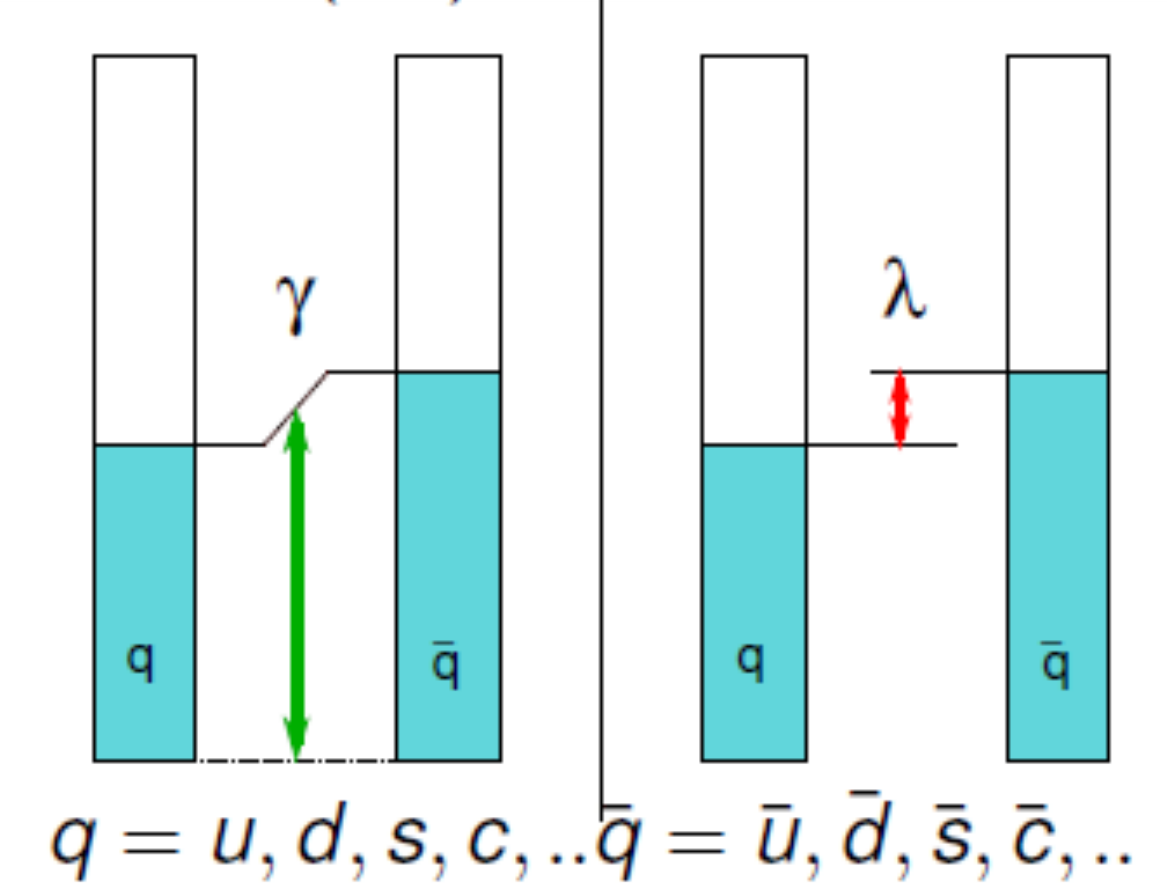}
\caption{Illustration of the difference in action of quantities required in the SHM for describing chemical conditions: the flavor conservation factor $\lambda$ and the flavor yield factor $\gamma$.}
\label{Fig2}
\end{figure}
\begin{itemize}
\item Flavor conservation factor $\lambda_q={\rm e}^{\mu/T}$:\\
it controls the difference between quarks and antiquarks of the same flavor $q-\bar{q}$, and describes ``relative" chemical equilibrium 
\item Flavor yield factor $\gamma_q$:\\
it measures the phase space occupancy, the absolute abundance of flavor $q$, controls the 
the number of quark-antiquark pairs $q+\bar{q}$, and describes ``absolute" chemical equilibrium
\item Overall fugacity $\Upsilon=\gamma\lambda$:\\
it is a product of the contributions from the constituent quark flavors contained in hadron $i$.
Example: $\Lambda(uds)$ is described with ($q=u,d$):\
$\Upsilon_{\Lambda(uds)}=\gamma_q^2\gamma_s\lambda_q^2\lambda_s$ 
and for the antilambda holds 
$\Upsilon_{\bar{\Lambda}(\bar{u}\bar{d}\bar{s})}=\gamma_q^2\gamma_s\lambda_q^{-2}\lambda_s^{-1}$ 
\item As noted above strangeness, (and charm) has similar fugacity factors as light quarks $q=u,d$ with flavor indicated by lower index.
\end{itemize}

The questions and topic we address in the following are: 
\begin{enumerate}
\item Does the SHM describe particle production at LHC, see Fig.~\ref{Fig3}? 
\begin{figure}[!ht]
\centering
\includegraphics[width=0.48\textwidth]{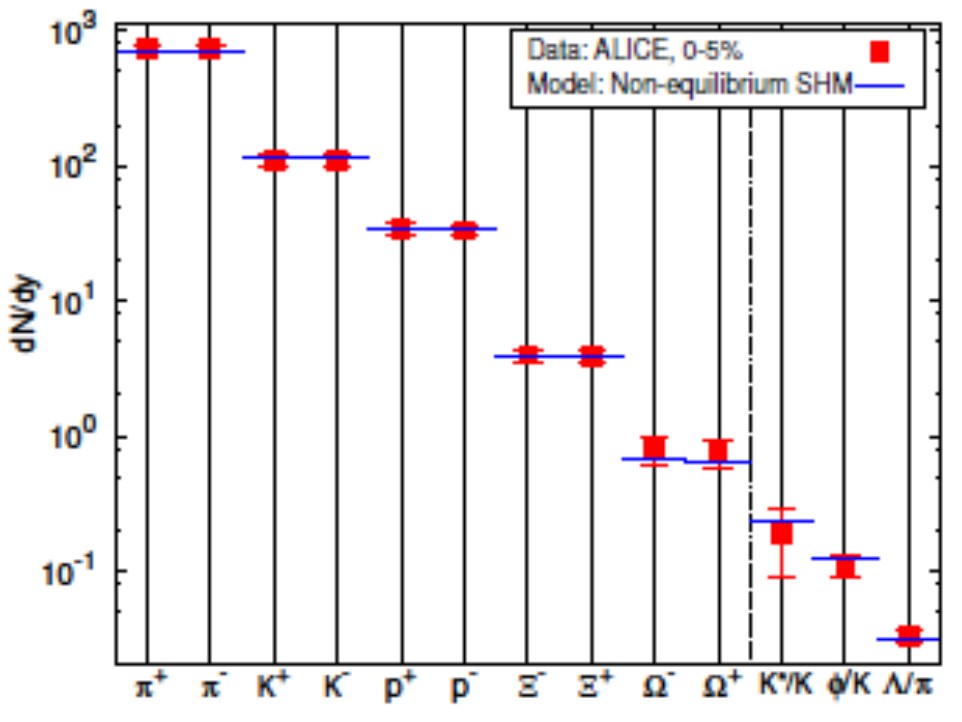}
\includegraphics[width=0.48\textwidth]{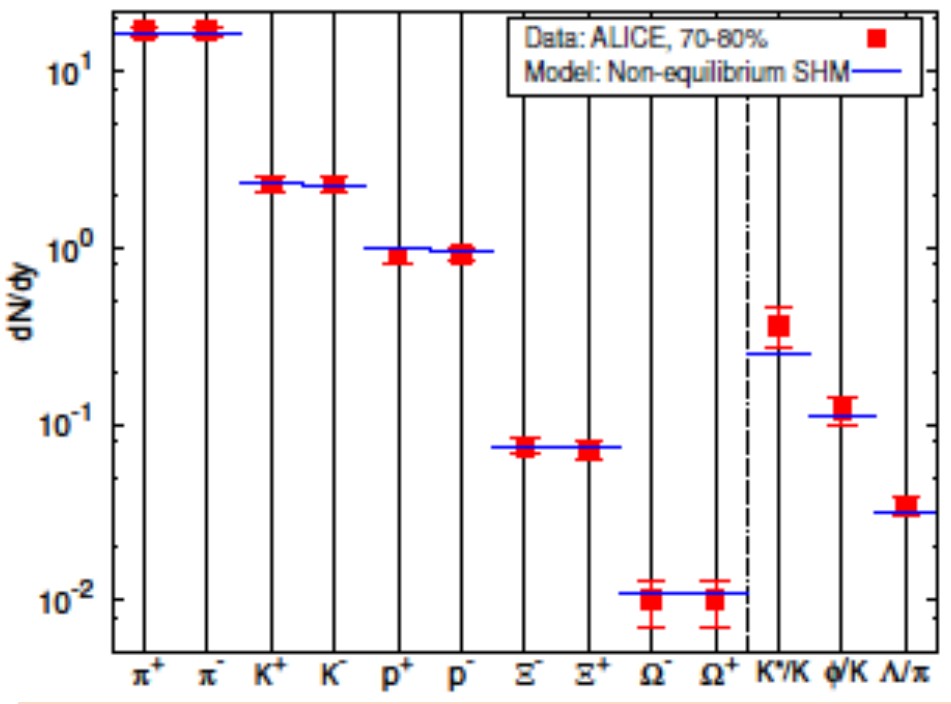}
\caption{Comparison of LHC-ALICE Collaboration results with the SHM: for both central (left panel) and peripheral (right panel) Pb-Pb collisions at $\sqrt{s_{NN}}=2.76$ TeV.}
\label{Fig3}
\end{figure}

\item How does the QGP fireball hadronize? 
\item What is same and what different comparing to RHIC
\item Universality across all collision energies: SPS, RHIC, LHC
\end{enumerate}

\section{SHM description of LHC results}
We test the SHM by applying it to the LHC data of the ALICE Collaboration obtained in the Pb-Pb collisions at $\sqrt{s_{NN}}=2.76$ TeV where the hadron yields span 5 orders of magnitude from central to peripheral collisions. The fit works perfectly, see Fig.~\ref{Fig3}, in the entire range of centralities. 

\begin{figure}[!ht]
\centering
\includegraphics[width=0.55\textwidth]{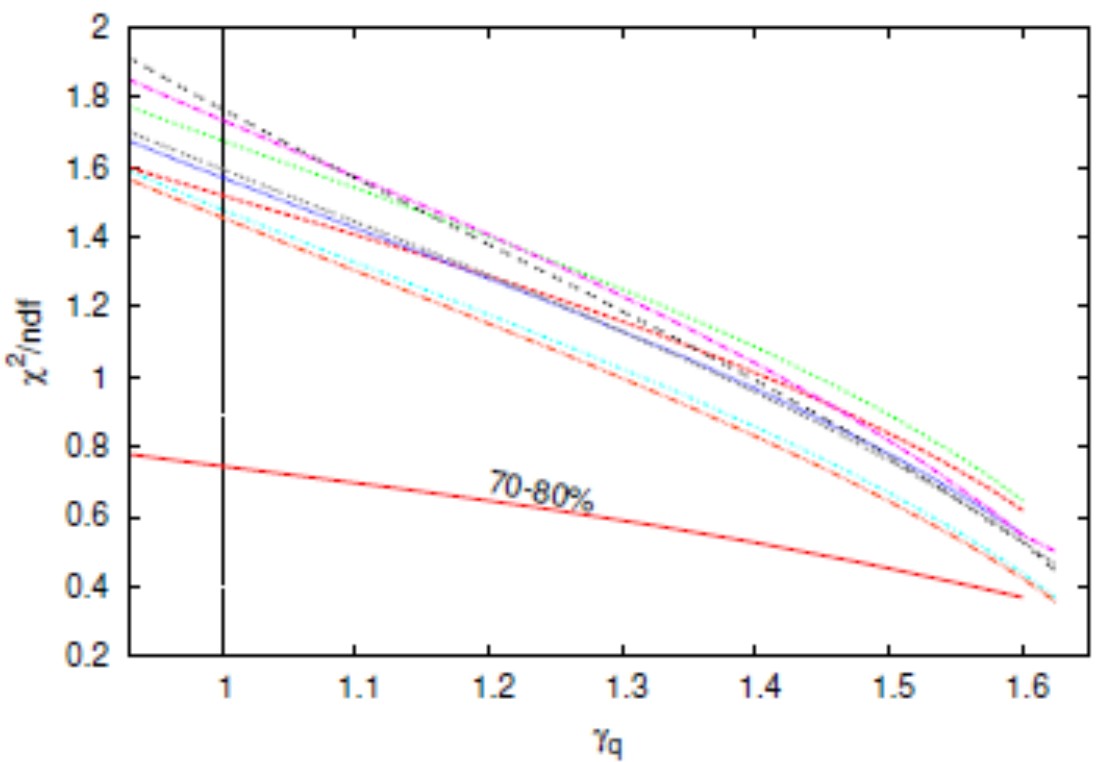}
\caption{The dependence of the $\chi^2$ for the fit on the flavor yield factor justifies the choice of $\gamma_q=1.6$ for best fit, see Ref.~\cite{Petran:2013lja}.}
\label{Fig4a}
\end{figure}

The salient element that makes the SHM work producing continuos and consistent results for the entropy content of the fireball is the allowance for chemical nonequilibrium of light quarks $\gamma_q\simeq 1.6$, an approach motivated by the fast hadronization hypothesis. The preference for this nonequilibrium value is seen for all centralities in Fig.~\ref{Fig4a} showing the profile of $\chi^2$ as a function of $\gamma_q$.

\begin{figure}[!ht]
\centering
\includegraphics[width=0.65\textwidth]{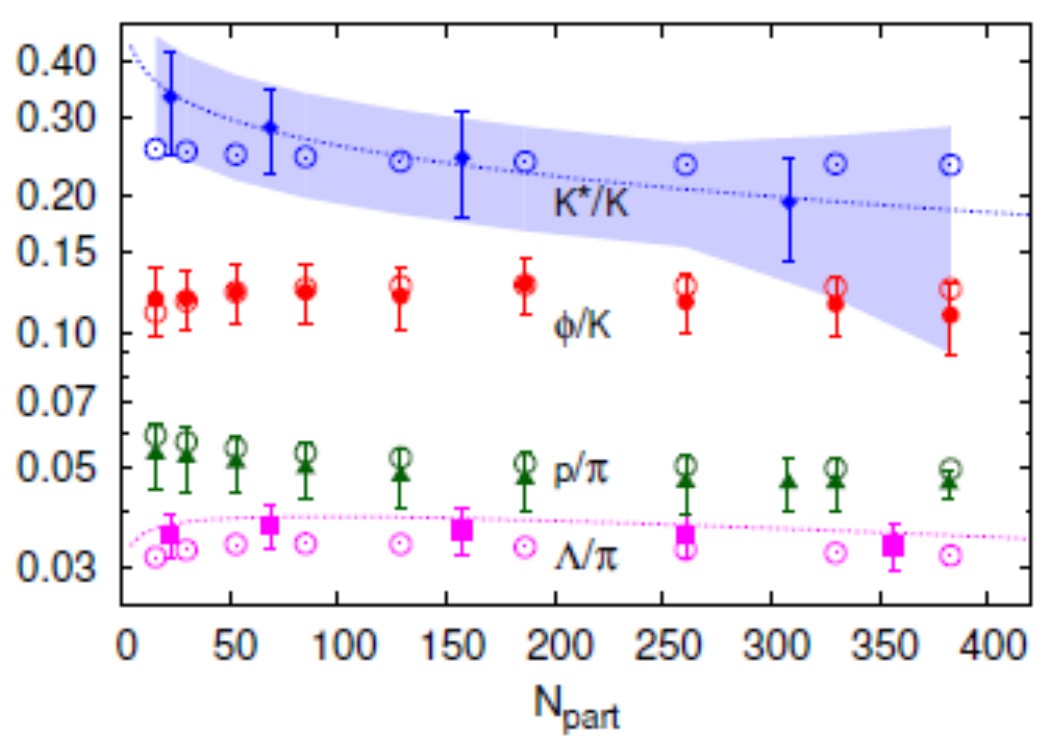}
\caption{Comparison of selected ratios in the ALICE experiment (full symbols) to the nonequilibrium SHM (open symbols) for the Pb-Pb collisions at $\sqrt{s_{NN}}=2.76$ TeV, see Ref.~\cite{Petran:2013lja}.}
\label{Fig4b}
\end{figure}

In Fig.~\ref{Fig4b} we see the centrality dependence of some of the particle yield ratios which play an important role in recognizing the chemical properties of the hadron source. Only with chemical nonequilibrium of light quarks, $\gamma_q=1.6$, we can describe all LHC data. No other approach works; the so-called ``afterburners" which can fix one particle yield, ruin the centrality systematics of other particles. 

\section{SHM description of RHIC-62}
Seen the success of our approach at LHC, we turn now to the question, does the SHM approach also describe particle production at RHIC? Clearly this question is a very large one and we focus here on one of the collision energies that we were able to look at comprehensively. This is the case of Au-Au results obtained at $\sqrt{s_{NN}}=62$ GeV. As the energy is lower, one expects that more peripheral collisions are less likely to reach a thermally equilibrated QGP stage. Thus another question arises: For how small a system is the hadronization at RHIC of same universal character we see it at LHC?

\begin{figure}[!ht]
\centering
\includegraphics[width=0.55\textwidth]{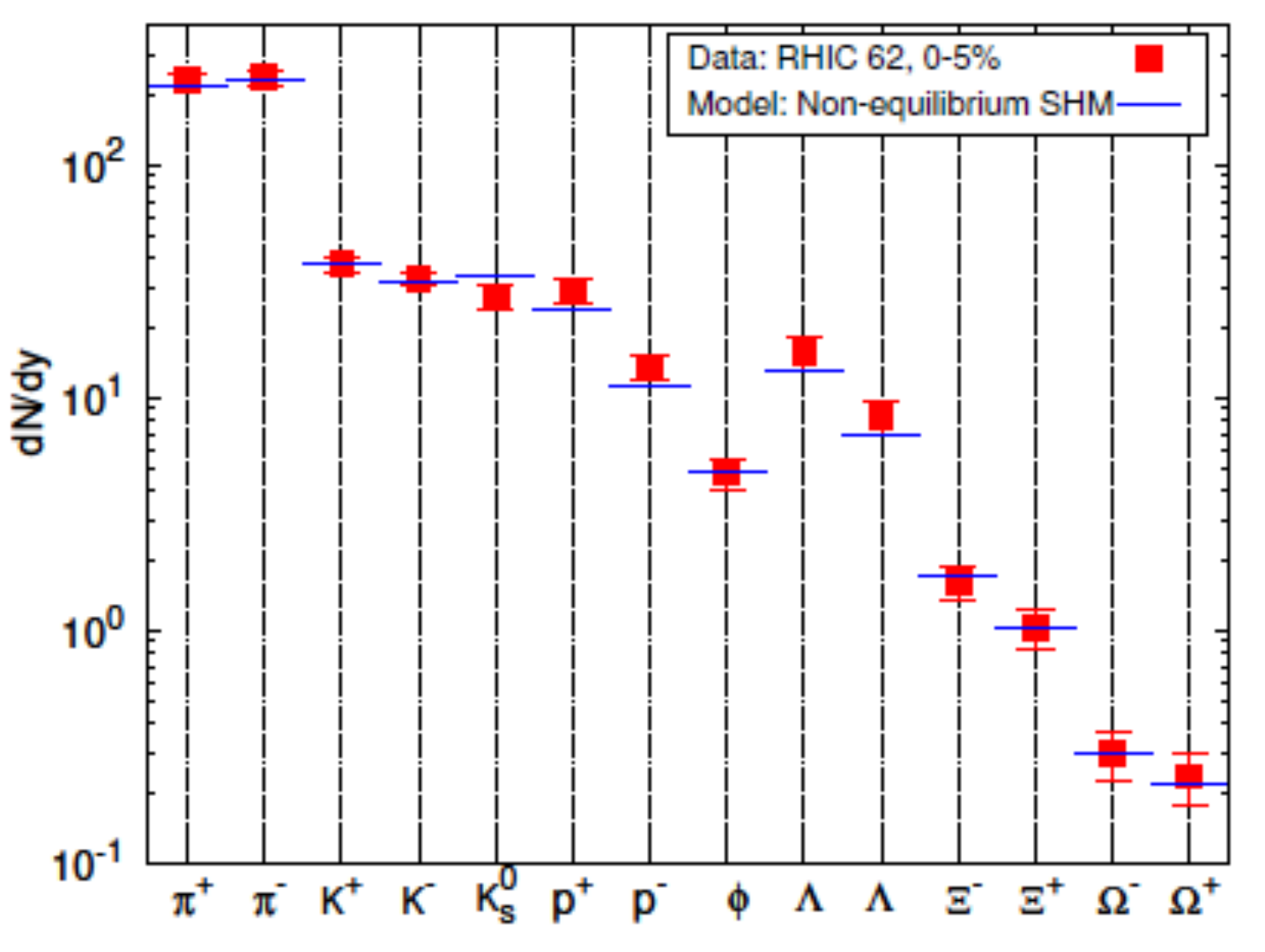}
\caption{The SHM analysis with SHARE (blue lines) of hadron yields from Au-Au collisions at RHIC for $\sqrt{s_{NN}}=62$ GeV (red symbols with error bars) results in a fit of freeze-out parameters $T=140$ MeV and $\mu_B=62.8$ MeV.}
\label{Fig5}
\end{figure}
We performed the analysis with SHARE for Au-Au collisions at RHIC $\sqrt{s_{NN}}=62$ GeV and have compared the results for the fireball properties with the LHC fits. 
The results are shown in Fig.~\ref{Fig5}, where an excellent fit ($\chi^2=0.38$) of STAR data 
\cite{Abelev:2008ab,Abelev:2008aa} is obtained with the model parameters: $T=140$ MeV, $dV/dy=850$ fm$^3$, $\gamma_q=1.6$, $\gamma_s=2.2$, $\lambda_q=1.16$, $\lambda_s=1.05$ corresponding to $\mu_B=62.8$ MeV \cite{Petran:2011aa}.

Having demonstrated that SHARE works very well in explaining hadron yields over the whole energy range between RHIC and LHC, we are well equipped for a discussion of similarities and differences.
We are particularly interested in strangeness as a signature of the QGP.

\section{Synthesis: LHC+RHIC+SPS}

The topic of a unified description of hadron production from SPS to RHIC, to LHC is discussed in Refs.~\cite{Rafelski:2014fqa,Rafelski:2009jr,Rafelski:2009gu}. For the physical properties of the fireball at freeze-out we find the energy density $\varepsilon=0.5$ GeV/fm$^3$, the pressure of $P=82$ MeV/fm$^3$ and the entropy density of $\sigma=3.3$ fm$^{-3}$.

\begin{figure}[!ht]
\centering
\includegraphics[width=0.78\textwidth]{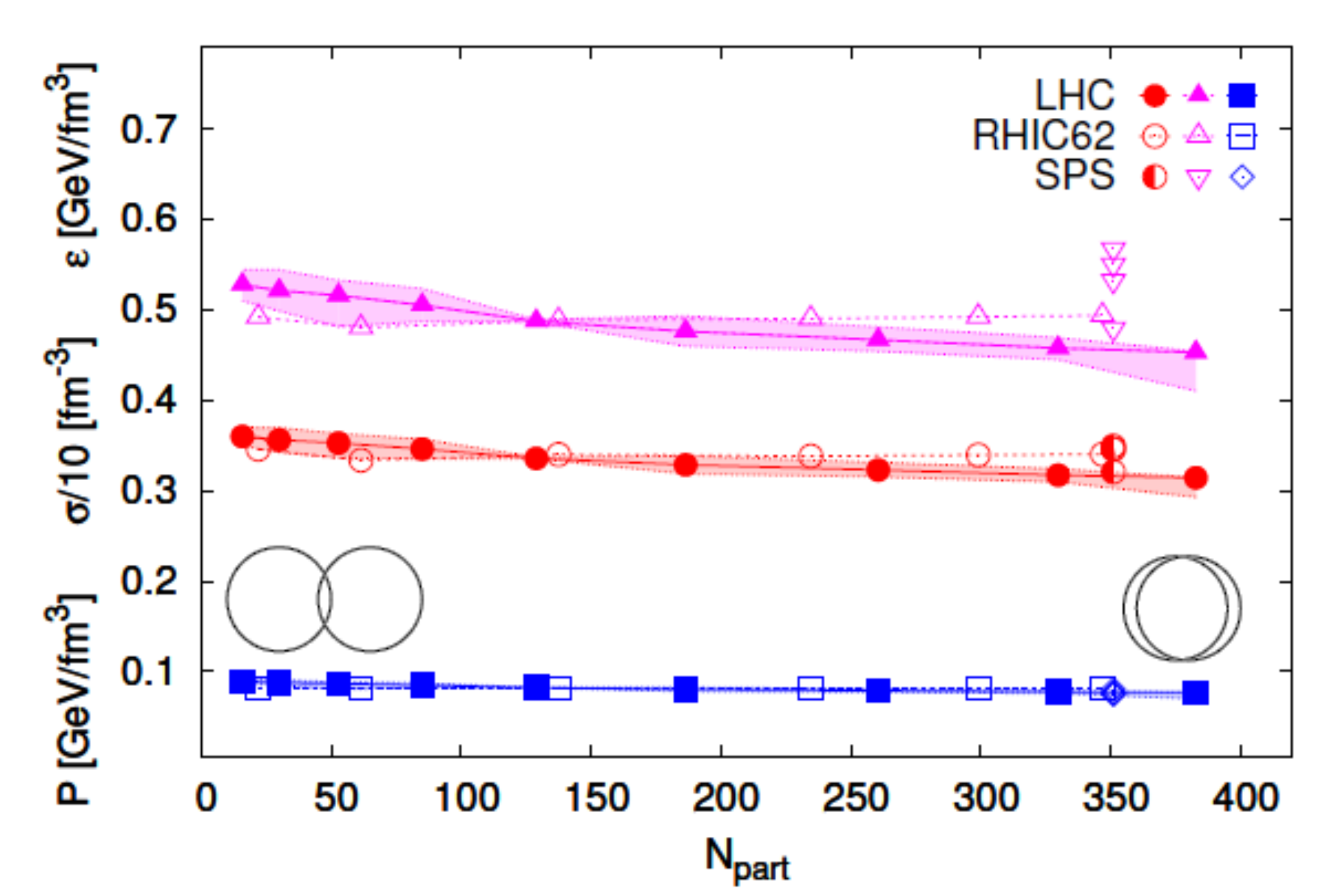}
\caption{Thermodynamic properties (pressure $P$, entropy density $\sigma$ and energy density $\varepsilon$) at freeze-out are universal \cite{Petran:2013qla}, nearly independent of collision energy and centrality.}
\label{Fig6}
\end{figure}

\begin{figure}[ht]
\centering
\includegraphics[width=0.48\textwidth]{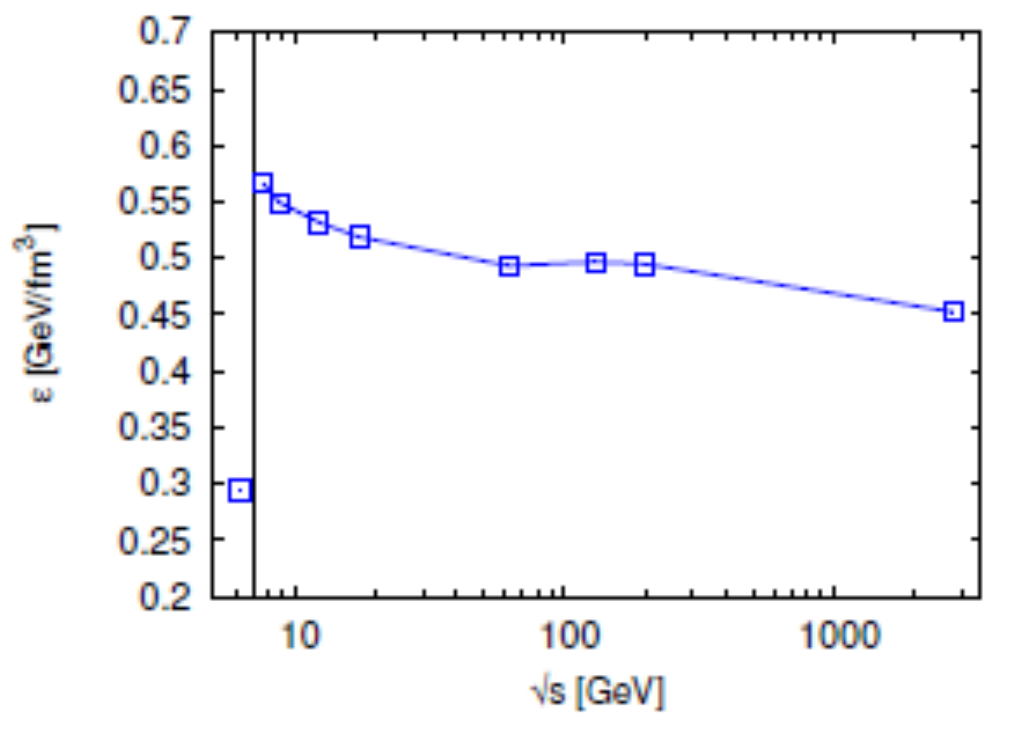}
\includegraphics[width=0.48\textwidth]{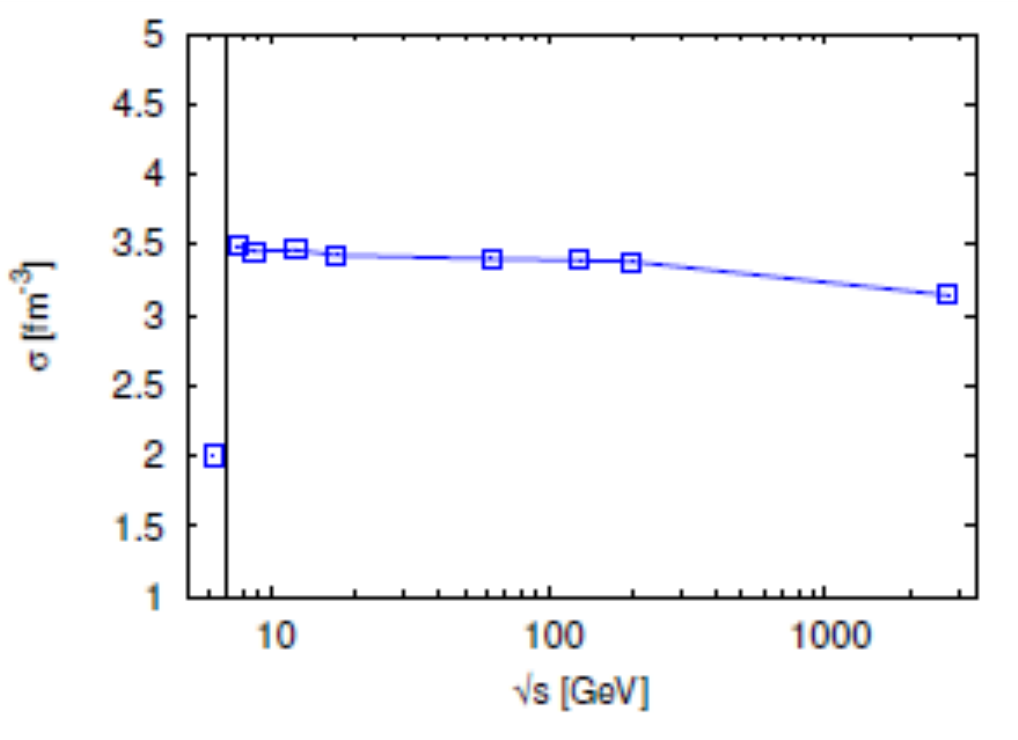}\\
\includegraphics[width=0.48\textwidth]{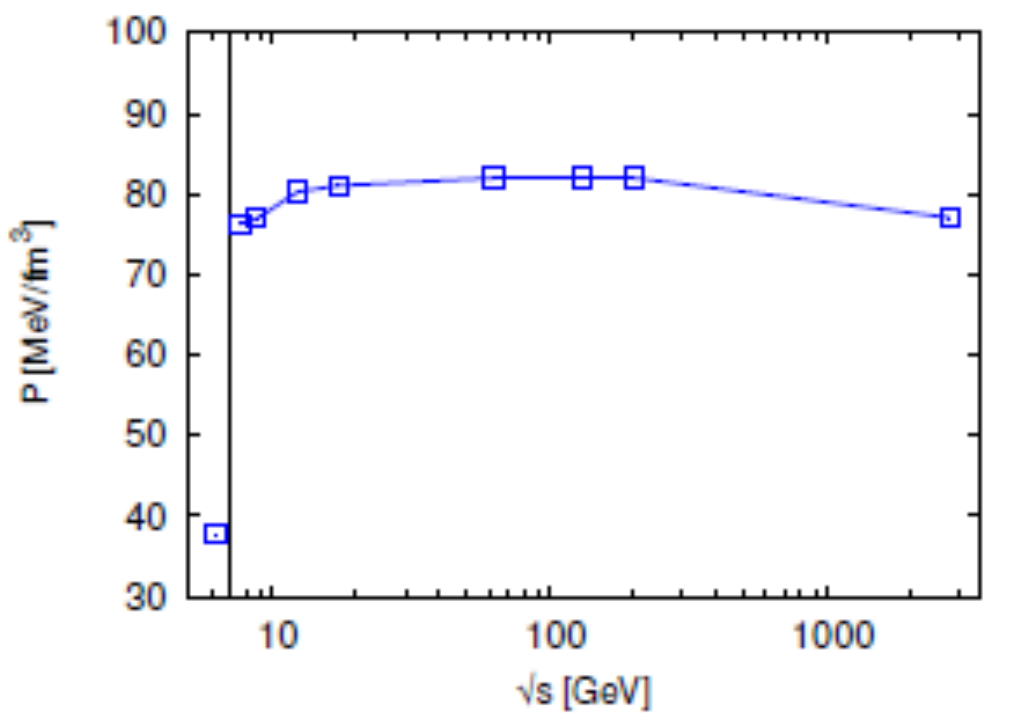}
\caption{Fireball bulk properties at freeze-out across the energies from SPS to RHIC, to LHC.}
\label{Fig11}
\end{figure}

In Fig.~\ref{Fig6} we show freeze-out bulk properties (pressure, entropy density and energy density) as a function of collision centrality at LHC and in Fig.~\ref{Fig11} the same for most central collisions as function of energy. These results demonstrates universality of freeze-out \cite{Petran:2013qla}, i.e., independence of collision energy and centrality.

\begin{figure}[!ht]
\centering
\includegraphics[width=0.95\textwidth]{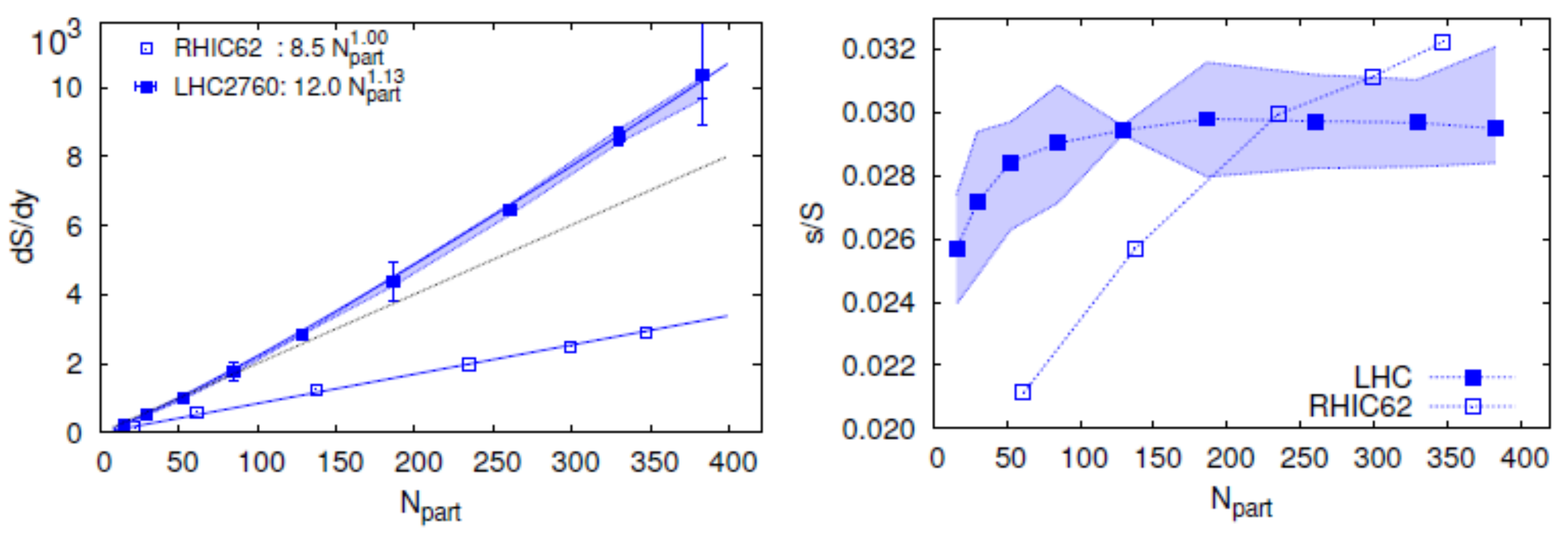}
\caption{The LHC-RHIC difference as a function of centrality: total entropy (left) and strangeness production per entropy $s/S$ (right), from Ref.~\cite{Petran:2013lja}.}
\label{Fig7}
\end{figure}

In Fig.~\ref{Fig7} we highlight the differences we found between LHC and RHIC that arise in the SHM interpretation of the particle yields. The centrality dependence of the total entropy (left panel) shows at LHC a steeper than linear behavior and an additional centrality dependent entropy production. The strangeness per entropy (right panel) shows a steeper increase at low $N_{part}$ and a quick saturation at a steady level for $N_{part}> 100$.

\begin{figure}[!ht]
\centering
\includegraphics[width=0.6\textwidth]{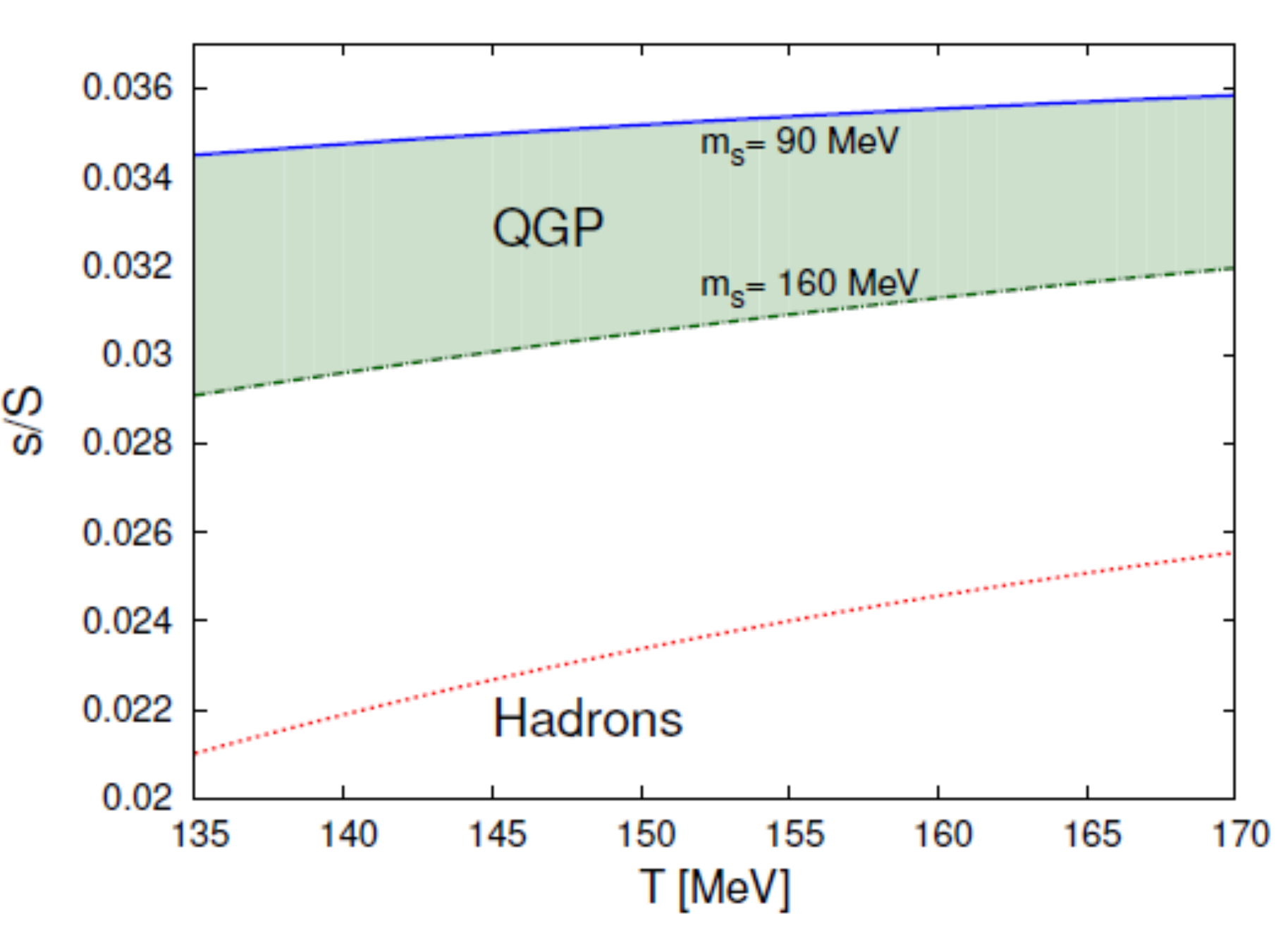}
\caption{Strangeness per entropy as a function of temperature in the hadron resonance gas and in the QGP.}
\label{Fig8}
\end{figure}

In Fig.~\ref{Fig8} we show the expected theoretical strangeness per entropy ratio~\cite{Kuznetsova:2006bh} for both QGP and hadron phases of matter, the hadron result was obtained for equilibrated HG using SHARE program. This theoretical result can be compared with the right panel in Fig.~\ref{Fig7}. We see that the experimental yields which emerged from our analysis are in the QGP domain for all LHC centralities. 

\begin{figure}[!ht]
\centering
\includegraphics[width=0.5\textwidth]{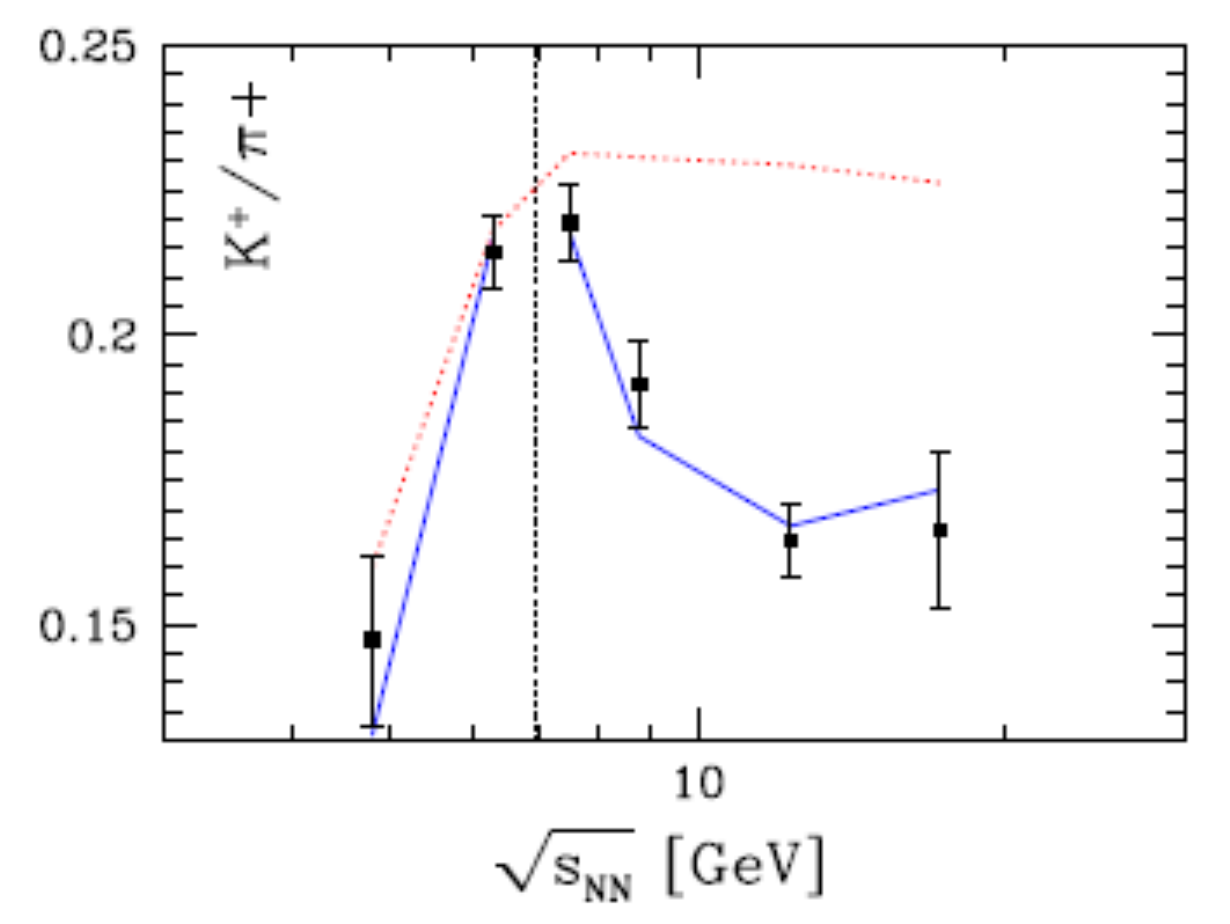}
\caption{The energy dependence of the $K^+/\pi^+$ ratio (``Marek's horn") is tracked perfectly with SHARE. Dotted line shows that not all models are able to agree with data.}
\label{Fig9}
\end{figure}

In Fig.~\ref{Fig9} it is shown~\cite{Rafelski:2009gu} that the peak in the beam energy dependence of the $K^+/\pi^+$ ratio (``Marek's horn") is tracked well as a function of energy with the nonequilibrium SHM fit provided by SHARE. 

\begin{figure}[!ht]
\centering
\hspace*{-0.8cm}\includegraphics[width=0.75\textwidth]{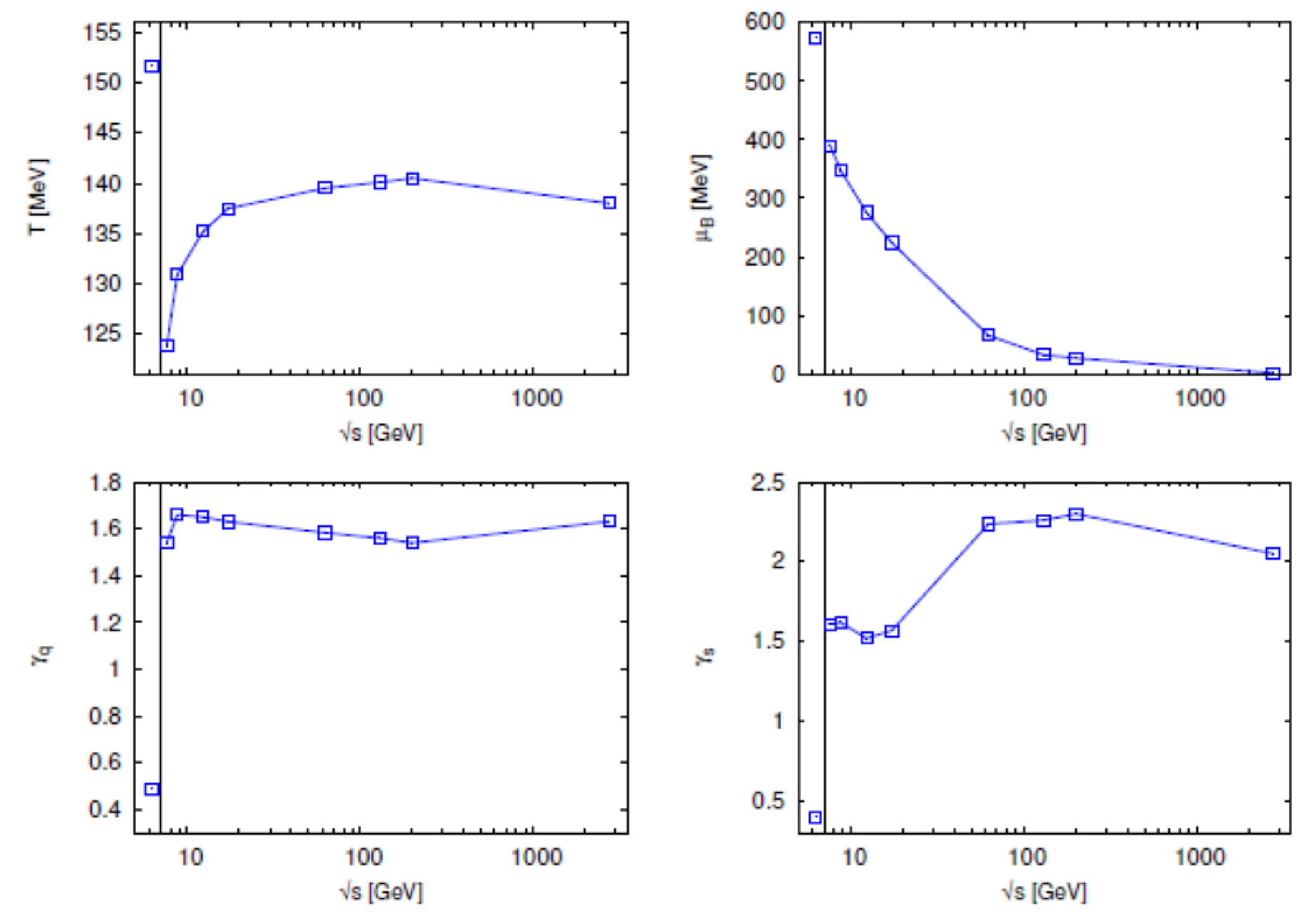}
\caption{Comparison of SHM parameters across the energies from SPS to RHIC, to LHC.}
\label{Fig10}
\end{figure}

The beam energy dependence of the SHM parameters are shown in Fig.~\ref{Fig10}, while 

\begin{figure}[ht]
\centering
\includegraphics[width=0.85\textwidth]{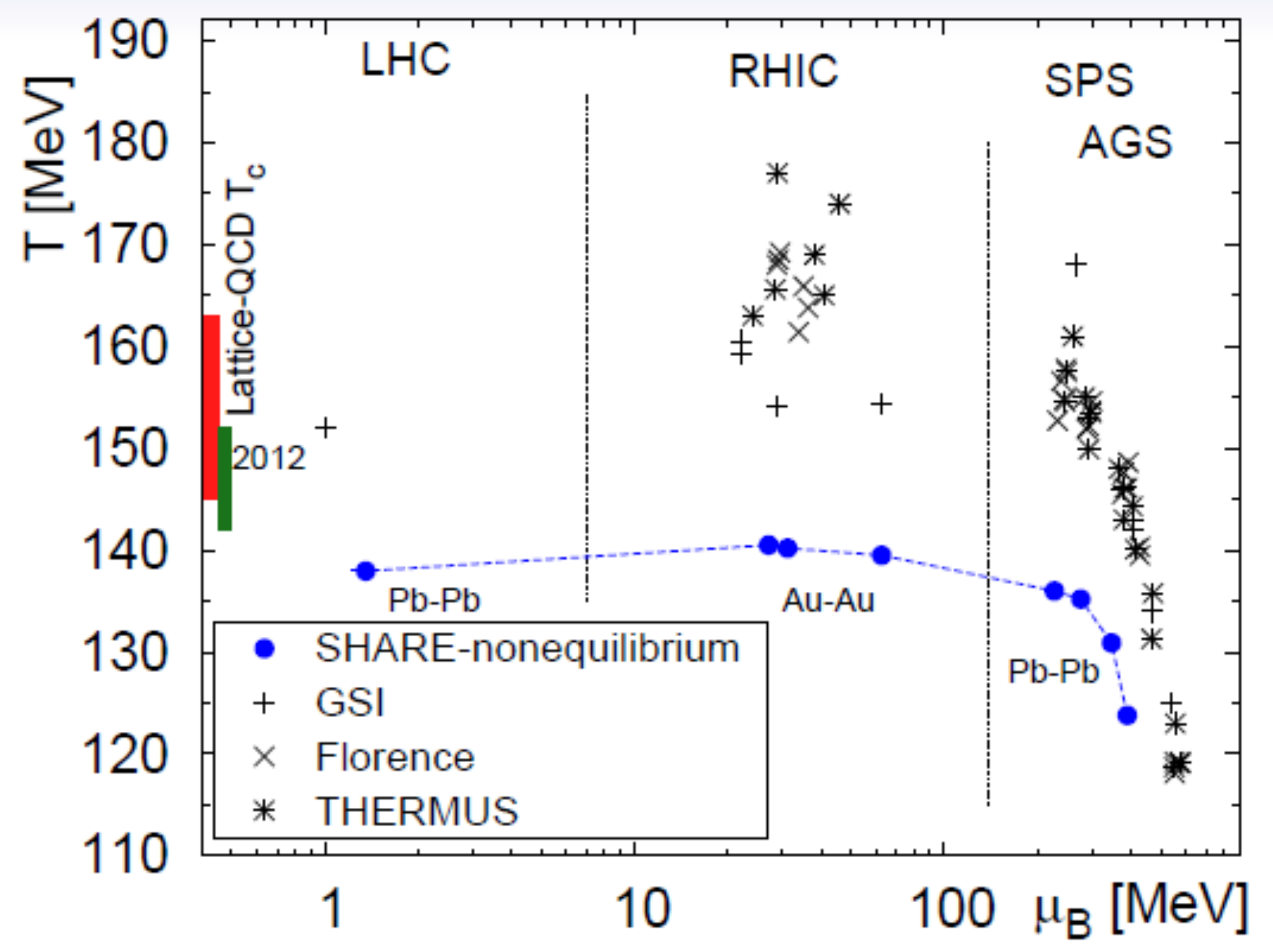}
\caption{Comparison of freeze-out parameters for different models with the pseudo-critical temperature from Lattice QCD \cite{Bazavov:2011nk}. Figure from Ref.~\cite{Petran:2013lja}.}
\label{Fig12}
\end{figure}

In Fig.~\ref{Fig12} we show a comparison of freeze-out parameters for different models with the pseudo-critical temperature from Lattice QCD \cite{Bazavov:2011nk}. For consistency reasons, the observed chemical freeze-out MUST be in the hadron resonance domain, i.e. below the lattice results for the pseudo-critical temperature of the chiral/deconfinement crossover transition. Clearly the recent lattice results rule out most hadronization models. We observe that since 1998 \cite{Letessier:1998sz} we have proposed and defended a view of hadronization that produces results fully consistent with present day lattice results.

\section{Conclusions}

Our exploration of phases of QCD matter relies on a precise method of hadron abundance analysis within the SHARE statistical hadronization model. Bulk properties of the QGP fireball are derived from physical properties of emitted hadronic particles. Hadronization of a relatively small and rapidly expanding drop of QGP considered in this work is not to be confounded with the lattice-QCD derived properties of infinite in size deconfined matter filling the early Universe. 

This study does not tell how a common QCD phase - the QGP state - was created at LHC, RHIC, {\it and} SPS, and how it evolves to hadronization. These paths could differ. However, we observe in the final state the same physical conditions of the fireball particle source - with varying hadronization volume $V$ or equivalently, total entropy content $S$ and strangeness $s$. In most central (head-on) high energy LHC collisions we observe relatively constant ratio $s/S$ indicating a QGP fireball near to full thermal equilibrium condition.

Given universal hadronization conditions that we have obtained we believe that when the QGP hadronizes it evaporates into free-streaming hadrons. Our analysis works without an interacting `phase' of hadrons. This is so since the use of light quark in QGP abundance, thus light quark overabundance among hadrons regulates just in correct way the observed hadron yields. This favors very good data fit since requilibration reactions following hadronization diminish agreement with production yields of multi-strange baryon and antibaryon.

\subsection*{Acknowledgments}

\begin{figure}[!ht]
\centering
\includegraphics[width=0.80\textwidth]{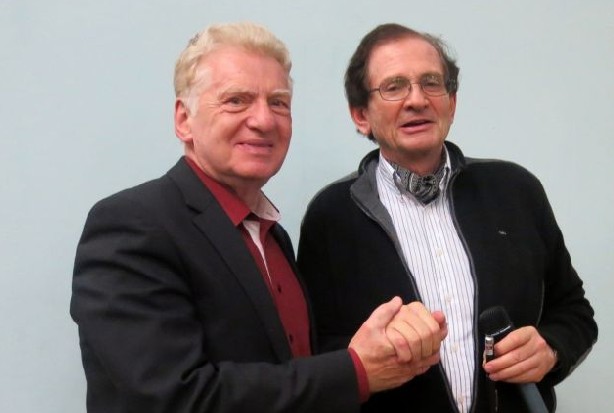}
\caption{Ludwik and Jan at the MB32 Symposium}
\label{Ludwik}
\end{figure}

We acknowledge the contributions by Inga Kuznetsova, Jean Letessier, Vojtech Petracek, and Giorgio Torrieri to the models and results presented here. Jan Rafelski acknowledges a long term collaboration with Ludwik Turko to whom this work is dedicated, see Fig.~\ref{Ludwik}.

\newpage

\end{document}